\documentclass[aps,prl,twocolumn,superscriptaddress]{revtex4-1}

\usepackage{amsfonts}
\usepackage{graphicx}
\usepackage{amsmath}
\usepackage{bm}
\usepackage[dvips]{color}
\usepackage{xcolor}

\newcommand{\be}{\begin{equation}}
\newcommand{\ee}{\end{equation}}
\newcommand{\bwt}{\begin{widetext}}
\newcommand{\ewt}{\end{widetext}}
\newcommand{\bea}{\begin{eqnarray}}
\newcommand{\eea}{\end{eqnarray}}
\newcommand{\ket}[1]{|#1\rangle}
\newcommand{\bra}[1]{\langle #1|}
\usepackage{verbatim}
\definecolor{purple}{RGB}{160,32,240}
\newcommand{\TK}[1]{{\bf #1}}

\newcommand{\degree}{\ensuremath{^\circ}}

\begin{document}
\title{Persistence of Hardy's nonlocality in time  }

\author{Sujit K Choudhary}
\email{choudhary@ukzn.ac.za}
\affiliation{School of Chemistry and Physics, University of KwaZulu--Natal, Private Bag X54001, 4000 Durban, South Africa.}
\author{Sandeep K Goyal}             
\email{goyal@ukzn.ac.za}
\affiliation{School of Chemistry and Physics, University of KwaZulu--Natal, Private Bag X54001, 4000 Durban, South Africa.}

\author{Thomas Konrad}
\email{konradt@ukzn.ac.za}
\affiliation{School of Chemistry and Physics, University of KwaZulu--Natal, Private Bag X54001, 4000 Durban, South Africa.}
\author{Sibasish Ghosh}
\email{sibasish@imsc.res.in}
\affiliation{Optics and Quantum Information Group, The Institute of Mathematical Sciences, 
CIT Campus, Taramani, Chennai 600113, India.}

\begin{abstract}
Hardy's nonlocality argument, which establishes incompatibility of quantum theory with 
local-realism, can also be used to reveal the time-nonlocal feature of quantum states.
For spin-$\frac{1}{2}$ systems, the  maximum probability of success of this argument is 
known to be $25\%$. We show that this maximum remains $25\%$ for all finite-dimensional 
quantum systems with suitably chosen observables. This enables a test of the quantum 
properties of macroscopic systems in analogy to the method of Leggett and Garg.
\end{abstract}

\maketitle
\section{INTRODUCTION}
 For testing the existence of superposition of macroscopically distinct
quantum states, Leggett and Garg \cite{legett} put forward the notion of {\it macrorealism}.
This notion rests on the classical paradigm \cite{brukner1, usha} that (i) physical properties of a
macroscopic object exist independent of the act of observation and (ii) any observable can be 
measured non-invasively, i.e., the ideal measurement of an observable at any instant of time does
not influence its subsequent evolution.

These original assumptions of \cite{legett}, namely the assumptions of  `macroscopic realism' and 
`noninvasive measurabilty', have been generalized to derive a temporal version of the Bell-CHSH  
inequality irrespective of whether the system under consideration is macroscopic or not 
\cite{brukner, generalized}.
Unlike the original Bell-CHSH scenario \cite{chsh} where correlations between measurement results from two distantly located physical systems are considered, temporal Bell-CHSH inequalities (or its generalizations)
are derived by focusing on one and the same physical system and analyzing the correlations
between measurement outcomes at two different times. These derivations are based on the following two assumptions: 
(i) {\it Realism}: The measurement results are determined by (possibly hidden) properties, which the particles
carry prior to and independent of observation, and (ii) {\it Locality in time}: The result 
of a measurement performed at time $t_2$ is independent of any  ideal measurement performed at some earlier
or later time $t_1$ \cite{tk}.

These inequalities get violated in Quantum Mechanics and thereby give rise to the notion of
{\it entanglement in time} which has been a topic of current research interest 
\cite {brukner,usha,ali,fritz,joag,white,brukner2}.
Interestingly, the original argument of Hardy, which establishes the incompatibility of 
Quantum Theory
with the notion of local-realism \cite{Hardy,hardy93}, can also be used to reveal this 
time-nonlocal feature of 
quantum states \cite{ali,fritz, white}. Recently, Hardy's argument was studied in the 
case of two observable settings at each time of measurement \cite{ali,fritz}. It was shown there that the maximum probability of success 
of this argument assumes  $25\%$ for a spin-$\frac{1}{2}$ particle \cite{ali,fritz},
the experimental verification of the above fact followed soon after in \cite{white}.

So far, only spin observables have been considered in the context of the temporal Hardy argument. 
We here study Hardy's argument for arbitrary observables of the system and find that the  
maximum success probability of this argument remains  $25\%$ irrespective of the dimension of 
the system. In addition, we argue that the same success probability can be observed with higher dimensional spin observables.  This is in sharp contrast with the findings of 
reference \cite{ali} where for spin observables it has been stated 
that the maximum probability of success of Hardy's argument decreases with increase
in spin value of the system involved. We also discuss the reason of this discrepancy.

\section{ Temporal version of nonlocality conditions for $d$-level systems }

Consider a single $d$ level physical system on 
which an observer (Alice) chooses to measure
one of two observables $\hat{A}_1$ or $\hat{A}_2$ at time $t_1$, whereas at a later time $t_2$,
another observer (Bob) \cite{bob} measures either of the two observables  $\hat{B}_1$ and $\hat{B}_2$. 
Let us refer  to an experiment with this setting in what follows as a "Hardy experiment".
Consider now the following set of conditions on the probabilities for Alice and
Bob to obtain outcomes $a_i$ and $b_j$ when measuring observables $\hat{A}_i$ and $\hat{B}_j$ 
respectively; $i$, $j\in \{1,2\}$ \cite{sk}:
\begin{align}
&{\rm prob}(\hat{A}_1,a_1~;\ \hat{B}_1,b_1)=0, \label{eqn-1}\\
&{\rm prob}(\hat{A}_1,\neg a_1~;\ \hat{B}_2,b_2)=0,\label{eqn-2}  \\
&{\rm prob}(\hat{A}_2,a_2~;\ \hat{B}_1,\neg b_1)=0, \label{eqn-3}\\
& {\rm prob}(\hat{A}_2,a_2~;\ \hat{B}_2,b_2) \ > 0 \label{eqn-4}.
\end{align}
The first condition says that if Alice chooses to measure the observable $\hat{A}_1$ and 
Bob  chooses observable $\hat{B}_1$, he will not obtain  $b_1$ as measurement result
whenever Alice has  detected the measurement value $a_1$. The remaining equations can be analyzed
in a similar manner ($\neg a_i$ denotes a measurement with any result other than $a_i$
and similarly $\neg b_j$ denotes a measurement with any result other than $b_j$).
These four conditions together form the basis of the temporal version of Hardy's
argument for $d$-level physical systems. This version of Hardy's argument makes use of the fact
that not all of the conditions (\ref{eqn-1})-(\ref{eqn-4}) can be simultaneously satisfied in a 
time-local realistic theory, but they can be in quantum mechanics.  

In a realistic theory, values are assigned to all the observables 
(whether or not they are actually measured) in such a manner that they 
agree with experimental observations. Consider a situation where a realist has been 
supplied with a table asking for the values of 
$\hat{A}_1$, $\hat{A}_2$, $\hat{B}_1$ and $\hat{B}_2$ in several runs of a Hardy experiment. 
In order to satisfy the last Hardy condition, he will have to assign $a_2$ for $\hat{A}_2$  and $b_2$ for $\hat{B}_2$ a few times. 
Out of these
few times,  he cannot choose neither values corresponding to  $\neg a_1$ for observable $\hat{A_1}$ nor values belonging to $\neg b_1$ for observable $\hat{B_1}$ since these events have zero probabilities according to conditions (\ref{eqn-2}) and  (\ref{eqn-3}), respectively. However, the alternative values these observables can assume, i.e.,  $ a_1$ and 
$b_1$ ,  also lead to a zero probability according to the first Hardy condition (\ref{eqn-1}). Hence in a realistic theory, 
for any choice of values of the observables $\hat{A_1}$ and $\hat{B_1}$ satisfying conditions (\ref{eqn-2})-(\ref{eqn-4}), 
the probability to obtain  $a_1$ for $\hat{A}_1$  and $b_1$ for $\hat{B}_1$ cannot be vanishing in contradiction with condition  
(\ref{eqn-1}). This is different for example in Quantum Mechanics where the probability distribution depends on the choice 
of observables  if this choice includes operators that do not commute.  Therefore, these conditions allow to distinguish 
whether a system can be described by a realistic theory or not. 

\section{Satisfaction of Temporal nonlocality conditions in quantum mechanics}

To show that the conditions (\ref{eqn-1})-(\ref{eqn-4}) can be simultaneously 
satisfied in Quantum Mechanics, we consider a quantum mechanical system in a pure state $|\psi\rangle$ with the associated 
Hilbert space $\mathcal{H}$. Moreover, we here restrict to projective measurements of observables $\hat{A}_i$ ($\hat{B}_i$)
that can be degenerate (so-called von Neumann- L\"{u}ders measurements \cite{lud}). A measurement of  $\hat{A}_i$ with result 
$a_i$ 
projects an initial state $\ket{\psi}$ onto the corresponding eigenspace specified by the projector 
\TK{$\Pi_{a_i,A_i}$,} i.e., the unnormalized state after projection reads $\ket{\psi'}=\Pi_{a_i,A_i}\ket{\psi}$, while the probability for this result to occur is given by $\langle\psi'|\psi'\rangle$ 
according to Born's rule. A measurement with any result other than $a_i$  projects onto a vector in the orthogonal 
complement given by the projector $\Pi_{\neg a_i, A_i}= \mathbb{I} -\Pi_{a_i, A_i}$ 
, where 
$\mathbb{I}$ denotes the identity operator
on Hilbert space $\mathcal{H}$.
Hence, Hardy's conditions (\ref{eqn-1})--(\ref{eqn-4}) can be expressed as expectation values of 
projectors as follows \cite{foot}: 
\small{
\begin{align}
&{\rm prob}(\hat{A}_1,a_1 ; \hat{B}_1,b_1)=\langle \psi
|\Pi_{a_1,{A_{1}}}\Pi_{b_1,{B_{1}}}\Pi_{a_1,{A_{1}}}|\psi\rangle=0,& \label{eqn-1a}\\
&{\rm prob}(\hat{A}_1,\neg a_1; \hat{B}_2,b_2)=\langle \psi
|\Pi_{\neg a_1,{A_{1}}}\Pi_{b_{2},{B_{2}}}\Pi_{\neg a_1,{A_{1}}}|\psi\rangle=0,&\label{eqn-2a}\\
&{\rm prob}(\hat{A}_2,a_2;  \hat{B}_1,\neg b_1)=\langle \psi
|\Pi_{a_2,{A_{2}}}\Pi_{\neg b_1,{B_{1}}}\Pi_{a_2,{A_{2}}}|\psi\rangle=0,& \label{eqn-3a}\\
&{\rm prob}(\hat{A}_2,a_2;  \hat{B}_2,b_2) =\langle \psi
|\Pi_{a_2,{A_{2}}}\Pi_{b_2,{B_{2}}}\Pi_{a_2,{A_{2}}}|\psi\rangle > 0. \label{eqn-4a}
\end{align}
}
\normalsize

These conditions lead to an upper bound of 1/4 for the last expression, independent of the dimension of 
the system as we will prove now. 

Equation (\ref{eqn-1a}) can be rewritten as 
\begin{equation}
 \langle \psi
|\Pi_{a_1,{A_{1}}}\Pi_{b_1,{B_{1}}}\Pi_{b_1,{B_{1}}}\Pi_{a_1,{A_{1}}}|\psi\rangle= 0\nonumber\
\end{equation}

This implies $\Pi_{b_1,{B_{1}}}\Pi_{a_1,{A_{1}}}|\psi\rangle= 0$ which further leads to    
 $\Pi_{a_1,{A_{1}}}|\psi\rangle=0$ or 
$\Pi_{b_1,{B_{1}}}\Pi_{a_1,{A_{1}}}|\psi\rangle=0$. 
Similarly from equation (\ref{eqn-2a}) one obtains $\Pi_{\neg a_1,{A_{1}}}|\psi\rangle=0$ or 
$\Pi_{b_2,{B_{2}}}\Pi_{\neg a_1,{A_{1}}}|\psi\rangle=0$.
Likewise equation (\ref{eqn-3a}) gives, $\Pi_{a_2,{A_{2}}}|\psi\rangle=0$ or 
$\Pi_{\neg b_1,{B_{1}}}\Pi_{a_2,{A_{2}}}|\psi\rangle=0$. But $\Pi_{a_2,{A_{2}}}|\psi\rangle=0$ contradicts the 
last of Hardy's conditions (\ref{eqn-4a}), so we discard it.

Hence for a quantum mechanical state $|\psi\rangle$ to exhibit Hardy's time-nonlocality, at least one of the following
sets of conditions must be simultaneously satisfied:
\begin{equation}
\label{cond1a}
\left.
\begin{array}{lcl}
\Pi_{a_1,{A_{1}}}|\psi\rangle=0,\Pi_{\neg a_1,{A_{1}}}|\psi\rangle=0,
\Pi_{\neg b_1,{B_{1}}}\Pi_{a_2,{A_{2}}}|\psi\rangle=0,\\
\langle \psi|\Pi_{a_2,{A_{2}}}\Pi_{b_2,{B_{2}}}\Pi_{a_2,{A_{2}}}|\psi\rangle > 0;
\end{array}
\right \}
\end{equation}
\begin{equation}
\label{cond2a}
\left.
\begin{array}{lcl}
\Pi_{a_1,{A_{1}}}|\psi\rangle=0,\Pi_{b_2,{B_{2}}}\Pi_{\neg a_1,{A_{1}}}|\psi\rangle=0,\\
\Pi_{\neg b_1,{B_{1}}}\Pi_{a_2,{A_{2}}}|\psi\rangle=0,
\langle \psi|\Pi_{a_2,{A_{2}}}\Pi_{b_2,{B_{2}}}\Pi_{a_2,{A_{2}}}|\psi\rangle > 0;
\end{array}
\right \}
\end{equation}
\begin{equation}
\label{cond3a}
\left.
\begin{array}{lcl}
\Pi_{b_1,{B_{1}}}\Pi_{a_1,{A_{1}}}|\psi\rangle=0, \Pi_{\neg a_1,{A_{1}}}|\psi\rangle=0,\\
\Pi_{\neg b_1,{B_{1}}}\Pi_{a_2,{A_{2}}}|\psi\rangle=0,
\langle \psi|\Pi_{a_2,{A_{2}}}\Pi_{b_2,{B_{2}}}\Pi_{a_2,{A_{2}}}|\psi\rangle > 0;
\end{array}
\right \}
\end{equation}

\begin{equation}
\label{cond4a}
\left.
\begin{array}{lcl}
\Pi_{b_1,{B_{1}}}\Pi_{a_1,{A_{1}}}|\psi\rangle=0, \Pi_{b_2,{B_{2}}}\Pi_{\neg a_1,{A_{1}}}|\psi\rangle=0,\\
\Pi_{\neg b_1,{B_{1}}}\Pi_{a_2,{A_{2}}}|\psi\rangle=0,
\langle \psi|\Pi_{a_2,{A_{2}}}\Pi_{b_2,{B_{2}}}\Pi_{a_2,{A_{2}}}|\psi\rangle > 0.
\end{array}
\right \}
\end{equation}
  
We continue by demonstrating that conditions (\ref{cond1a}), (\ref{cond3a}) and (\ref{cond4a}) are inconsistent in themselves
while conditions (\ref{cond2a}) imply a maximal success probability of 1/4 as claimed above. \\

The first two conditions
of (\ref{cond1a}), namely, $\Pi_{a_1,{A_{1}}}|\psi\rangle=0$ and $\Pi_{\neg a_1, {A_{1}}}|\psi\rangle=0 $  
cannot be simultaneously true.

The first condition of (\ref{cond3a}), namely $\Pi_{b_1,{B_{1}}}\Pi_{a_1,{A_{1}}}|\psi\rangle=0$ 
implies that $\Pi_{a_1,{A_{1}}}|\psi\rangle$ is a vector in the 
orthogonal complement of the image of $\Pi_{b_1,B_1}$, i.e.,
\begin{equation}\label{11-1}
 \Pi_{a_1,{A_{1}}} \leq \Pi_{\neg b_1,{B_{1}}},
\end{equation}
defined by 
$\Pi_{\neg b_1, B_1} - \Pi_{a_1, A_1}$  being a positive operator.
Equation (\ref{11-1}) can be rewritten as  
\begin{equation}\label{11-2}
 \Pi_{b_1,{B_{1}}} \leq \Pi_{\neg a_1,{A_{1}}},
\end{equation}
Similarly, from the third condition of (\ref{cond3a}), namely from 
$\Pi_{\neg b_1,{B_{1}}}\Pi_{a_2,{A_{2}}}|\psi\rangle=0$, one obtains 
\begin{equation}\label{11-3}
 \Pi_{a_2,{A_{2}}} \leq \Pi_{b_1,{B_{1}}},
\end{equation}
Equations (\ref{11-2}) and (\ref{11-3}) together imply
\begin{equation}\label{11-4}
 \Pi_{a_2,{A_{2}}} \leq \Pi_{\neg a_1,{A_{1}}},
\end{equation}
and hence
\begin{equation}\label{11-5}
 \langle \psi|(\Pi_{a_2,{A_{2}}}-\Pi_{\neg a_1,{A_{1}}})|\psi\rangle\le 0.
\end{equation}
 Taking into account  the  second condition of (\ref{cond3a}), namely $\Pi_{\neg a_1,{A_{1}}}|\psi\rangle=0$ 
equation (\ref{11-5}) yields
\begin{equation}\label{11-6}
 \langle \psi|\Pi_{a_2,{A_{2}}}|\psi\rangle\le 0.
\end{equation}
But, $\langle \psi|\Pi_{a_2,{A_{2}}}|\psi\rangle$ is the probability of obtaining $a_2$ as
measurement result in a measurement of observable $\hat{A}_2$ on $|\psi\rangle$, so it can neither
be negative, nor can it  be equal to zero as this would imply that the probability in the last of Hardy's conditions (\ref{eqn-4a}) vanishes. 

Using a similar argument we show in Appendix A that the first three conditions of 
(\ref{cond4a}) imply $\langle \psi|\Pi_{a_2,{A_{2}}}
\Pi_{ b_2,{B_{2}}}\Pi_{a_2,{A_{2}}}|\psi\rangle\le0 $ which contradicts the last condition in 
(\ref{cond4a}).

Thus we are left with the set (\ref{cond2a}) only. From its first condition,
namely, from $\Pi_{a_1,{A_{1}}}|\psi\rangle=0$, it follows that
$\Pi_{\neg a_1,{A_{1}}}|\psi\rangle=|\psi\rangle$. The second condition of (\ref{cond2a}) then reads
\begin{equation}\nonumber
 \Pi_{b_2,{B_{2}}}|\psi\rangle=0
\end{equation}
 which implies
\begin{equation}\label{new1}
 |\psi\rangle\langle \psi|\leq\Pi_{\neg b_2,{B_{2}}};
\end{equation}
 Equation (\ref{new1}) can be rewritten as
\begin{equation}\label{new2}
 \Pi_{b_2,{B_{2}}}\leq\mathbb{I}-|\psi\rangle\langle \psi|
\end{equation}
The nonzero probability in Hardy's argument thus reads
\small{
\begin{align}
&\langle \psi|\Pi_{a_2,{A_{2}}}\Pi_{b_2,{B_{2}}}\Pi_{a_2,{A_{2}}}|\psi\rangle &\nonumber\\
&\leq  
\langle \psi|\Pi_{a_2,{A_{2}}} (\mathbb{I}-|\psi\rangle\langle \psi|)\Pi_{a_2,{A_{2}}}|\psi\rangle&\nonumber\\  
&=  \langle \psi|\Pi_{a_2,{A_{2}}}\Pi_{a_2,{A_{2}}}|\psi\rangle-
\langle \psi|\Pi_{a_2,{A_{2}}}|\psi\rangle
\langle \psi|\Pi_{ a_2,{A_{2}}}|\psi\rangle&\nonumber\\
&=   \langle \psi|\Pi_{a_2,{A_{2}}}|\psi\rangle - (\langle \psi|\Pi_{a_2,{A_{2}}}|\psi\rangle)^2&\nonumber\\
&\leq    \frac{1}{4}&\,\label{newin}
\end{align}
}
\normalsize
as the maximum value of 
$ \langle \psi|\Pi_{a_2,{A_{2}}}|\psi\rangle - 
(\langle \psi|\Pi_{a_2,{A_{2}}}|\psi\rangle)^2$ is 1/4 and in which case
$\langle \psi|\Pi_{a_2,{A_{2}}}|\psi\rangle=\frac{1}{2}$.\\
 The maximum probability is achieved, e.g., if  the measurements are chosen such that
\begin{equation}
\Pi_{\neg b_2,{B_{2}}} = \Pi_{\neg a_1,{A_{1}}} =\sum_{i=1}^n\ket{\psi_i}\bra{\psi_i} < \mathbb{I} \label{choice12}
\end{equation}
 for an orthonormal set of vectors $\ket{\psi_i}$ containing the initial state vector, say $\ket{\psi}=(\ket{\varphi}+\ket{\varphi^\perp})/\sqrt{2}=\ket{\psi_1}$  and 
\begin{equation}
\Pi_{ a_2,{A_{2}}} = \Pi_{ b_1,{B_{1}}}= \sum_{i=1}^m\ket{\varphi_i}\bra{\varphi_i}< \mathbb{I} 
\end{equation}
for an orthonormal set of vectors $\ket{\varphi_i}$ that span a subspace which comprises one component $\ket{\varphi}$ of the initial state but not the perpendicular one $\ket{\varphi^\perp}$. This guarantees  the third condition in (\ref{cond2a}) while the first two are satisfied merely due to the choice (\ref{choice12}). In addition, in order to yield the maximum probability (i.e., for equality in (\ref{newin}) to hold) the vectors $\ket{\psi_i}$ can be chosen such that  $ \bra{\psi_i}\varphi\rangle=0$ for $i=2, 3\ldots n$.

Thus the success probability of  Hardy's temporal nonlocality argument can go up to $25\%$ in quantum theory
irrespective of the dimension of the system.\\

This result differs from the result obtained in \cite{ali} because there apparently only a restricted
set of observables and states of spin systems were considered. For example, for spin $s=1$ (a three level
system) reference \cite{ali} claims a maximal success probability of (1/16). 
However, it can be checked (cf. Appendix-B) 
that for such a system, the following setting achieves a maximal success probability
 1/4 of Hardy's argument in agreement with the upper bound shown above: 
\begin{equation}\nonumber
 |\psi\rangle=\left( \begin{array}{cccc}
0 \\
 -\sin\alpha  \\
\cos\alpha\end{array} \right) 
\end{equation}

\begin{equation}\nonumber
\hat{A}_1=\hat{B}_2=\left(\begin{array}{ccc}
1 &0  & 0 \\
0 & 0 & 0 \\
0 & 0 & -1 \end{array} \right)=\hat{S}_Z
\end{equation}
\begin{equation}\nonumber
 \hat{A}_2=\hat{B}_1=\cos\alpha\hat{S}_Z-\sin\alpha\hat{S}_X
\end{equation}
where $\alpha=\cos^{-1}(\sqrt{2}-1), \mbox{i.e.,}~ \alpha\approx 65.53\degree$ and
\begin{equation}\nonumber
 \hat{S}_X=\left(\begin{array}{ccc}
0 &\frac{1}{\sqrt{2}}  & 0 \\
\frac{1}{\sqrt{2}} & 0 & \frac{1}{\sqrt{2}} \\
0 & \frac{1}{\sqrt{2}} & 0 \end{array} \right).
\end{equation}

From the set of conditions (\ref{cond2a}), it is clear that for a given 
observable setting, there can be more than one
pure state exhibiting Hardy's time-nonlocality. Hence a mixture of them will also exhibit this 
nonlocality. But, as the success probability of
Hardy's argument in this case is a convex sum of the success probabilities for individual 
pure states, hence for mixed states too, 
the maximum success probability of Hardy's argument cannot go beyond 1/4.

\section{conclusion}

In conclusion, we have shown that the maximum of the success probability appearing in the temporal
version of Hardy's argument
is $25\%$ irrespective of the dimension of the quantum mechanical system and the type of observables involved. 
For the special case of spin measurements, for spin-1 and spin-3/2 observables, we have shown in Appendix-B that this maximum
can also  be achieved. Moreover, we conjecture that this maximum  can be observed for any spin system. 
Thus this temporal nonlocality persists as opposed to the  idea that quantum systems with higher dimensional state space 
behave more classical which was put forward  in \cite{cliffton} for the case of spatial non locality. Our result is at par with the findings for spatially 
separated systems where the success probability for Hardy's argument is also independent of the dimension of systems' Hilbert space  
\cite{sibada, scarani}. Moreover, contrary to the implications from \cite{ali}, our result ensures the 
possibility to probe the existence of quantum superpositions for macroscopic systems by means of 
Hardy's argument and thus independent of the Leggett-Garg inequality \cite{legett}.  

We have given a recipe to achieve a maximal success probability for Hardy's argument for general
(degenerate or non-degenerate) observables. Note, however that each of them can be replaced
by a dichotomic, degenerate observable by only distinguishing the cases where a certain measurement result
(say $a$) occurs from the cases where it does not ($\neg a$), cp. conditions (\ref{eqn-1})-(\ref{eqn-4}).
This is in agreement with our result that the maximal success probability in $d$-dimensions is the same as for 
the qubit case which only features dichotomic observables. Thus the maximum value of the success probability
in Hardy's argument for a $d$-level system, will remain one and the same in the framework 
of all generalized time-nonlocal theories constrained only by `no signalling in time'  \cite{brukner3}
where a measurement does not change the outcome statistics of a later measurement.  

 \renewcommand{\theequation}{A-\arabic{equation}}
 \setcounter{equation}{0}  
\section{appendix A}
The first condition of (\ref{cond4a}) implies
\begin{equation}\label{12-1}
 \Pi_{a_1,{A_{1}}} \leq \Pi_{\neg b_1,{B_{1}}},
\end{equation}
which can be rewritten as
\begin{equation}\label{12-2}
 \Pi_{b_1,{B_{1}}} \leq \Pi_{\neg a_1,{A_{1}}}
\end{equation}
From the second condition of (\ref{cond4a}), it follows that
\begin{equation}\label{12-3}
 \Pi_{\neg a_1,{A_{1}}} \leq \Pi_{\neg b_2,{B_{2}}}
\end{equation}
Equations (\ref{12-2}) and (\ref{12-3}) together imply
\begin{equation}\label{12-4}
 \Pi_{ b_1,{B_{1}}} \leq \Pi_{\neg b_2,{B_{2}}}
\end{equation}
The third condition of (\ref{cond4a}) gives
\begin{equation}\label{12-5}
 \Pi_{ a_2,{A_{2}}} \leq \Pi_{b_1,{B_{1}}}
\end{equation}
From (\ref{12-4}) and (\ref{12-5}), one obtains 
\begin{equation}\label{12-6}
 \Pi_{ a_2,{A_{2}}} \leq \Pi_{\neg b_2,{B_{2}}},
\end{equation}
which gives
\small{
\begin{align}\allowbreak\label{12-7}
&\langle \psi|\Pi_{a_2,{A_{2}}}(\Pi_{\neg b_2,{B_{2}}}-
\Pi_{a_2,{A_{2}}})\Pi_{a_2,{A_{2}}}|\psi\rangle\ge 
0&\nonumber\\
&\Leftrightarrow \langle \psi|\Pi_{a_2,{A_{2}}}\Pi_{\neg b_2,{B_{2}}}\Pi_{a_2,{A_{2}}}|\psi\rangle-
\langle \psi|\Pi_{a_2,{A_{2}}}|\psi\rangle\ge 
0&\nonumber\\
&\Leftrightarrow  \langle \psi|\Pi_{a_2,{A_{2}}}|\psi\rangle-\langle \psi|\Pi_{a_2,{A_{2}}}
\Pi_{ b_2,{B_{2}}}\Pi_{a_2,{A_{2}}}|\psi\rangle\nonumber\\
&-\langle \psi|\Pi_{a_2,{A_{2}}}|\psi\rangle\ge 
0&\nonumber\\
&\Leftrightarrow  \langle \psi|\Pi_{a_2,{A_{2}}}
\Pi_{ b_2,{B_{2}}}\Pi_{a_2,{A_{2}}}|\psi\rangle\le0&\nonumber
\end{align}
}
\normalsize

\renewcommand{\theequation}{B-\arabic{equation}}
 \setcounter{equation}{0}  
\section{appendix B}

\subsection*{Case of spin-1 systems}
 Hardy's time-nonlocality conditions  for a 
three level system as given in \cite{ali}: 
\begin{equation}
\label{prob1}
\left.
\begin{array}{lcl}
\rm prob(\hat{A}_1=+1,\ \hat{B}_1=+1)=0, \\
\rm prob(\hat{A}_1=0,\ \hat{B}_2=+1)=0, \\
\rm prob(\hat{A}_1=-1,\ \hat{B}_2=+1)=0, \\
\rm prob(\hat{A}_2=+1,\ \hat{B}_1=0)=0, \\
\rm prob(\hat{A}_2=+1,\ \hat{B}_1=-1)=0, \\
\rm prob(\hat{A}_2=+1,\ \hat{B}_2=+1)=q \ > 0.
\end{array}
\right \}
\end{equation}
As mentioned in Section III, the following observable-state setting achieves maximum success 
probability 1/4 of Hardy's time-nonlocal argument:  
 \begin{equation}\nonumber
 |\psi\rangle=-\sin\alpha|\hat{S}_Z=0\rangle+\cos\alpha|\hat{S}_Z=-1\rangle
\end{equation}
\begin{equation}\nonumber
 \hat{A}_1=\hat{B}_2=\hat{S}_Z
\end{equation}
\begin{equation}\nonumber
 \hat{A}_2=\hat{B}_1=\cos\alpha\hat{S}_Z-\sin\alpha\hat{S}_X
\end{equation}
where $\alpha=\cos^{-1}(\sqrt{2}-1), \mbox{i.e.,}~ \alpha\approx 65.53\degree$,
\begin{equation}\nonumber
 \hat{S}_Z=\left(\begin{array}{ccc}
+1 &0  & 0 \\
0 & 0 & 0 \\
0 & 0 & -1 \end{array} \right),
\end{equation}
\begin{equation}\nonumber
 \hat{S}_X=\left(\begin{array}{ccc}
0 &\frac{1}{\sqrt{2}}  & 0 \\
\frac{1}{\sqrt{2}} & 0 & \frac{1}{\sqrt{2}} \\
0 & \frac{1}{\sqrt{2}} & 0 \end{array} \right)
\end{equation}
and $\Pi_{a_1,{A_{1}}},\Pi_{a_2,{A_{2}}}, \Pi_{b_1,{B_{1}}},\Pi_{b_2,{B_{2}}}$ are given as follows:
\begin{eqnarray}\nonumber
 \Pi_{a_1,{A_{1}}}=|\hat{S}_Z=+1\rangle\langle\hat{S}_Z=+1|=\Pi_{b_2,{B_{2}}}\nonumber\\
\Pi_{a_2,{A_{2}}}=|\hat{A}_2=+1\rangle\langle\hat{A}_2=+1|=\Pi_{b_1,{B_{1}}}\nonumber
\end{eqnarray}
where 
\begin{eqnarray}\nonumber
 &|\hat{A}_2=+1\rangle=\cos^{2}\frac{\alpha}{2}|\hat{S}_Z=+1\rangle&\nonumber\\
&-\frac{\sin\alpha}{\sqrt{2}}|\hat{S}_Z=0\rangle+
\sin^{2}\frac{\alpha}{2}|\hat{S}_Z=-1\rangle&\nonumber.
\end{eqnarray}
It can be checked that $\Pi_{+1,{A_{2}}}|\psi\rangle$ is of the form:
\begin{equation}\label{extra1}
 \Pi_{+1,{A_{2}}}|\psi\rangle=\frac{1}{2}|\psi\rangle+\frac{1}{2}|\phi\rangle,
\end{equation}
where $|\phi\rangle=|\hat{S}_Z=+1\rangle$, which is orthogoanl to $|\psi\rangle$.

\subsection*{Case of spin-3/2 systems}

Hardy's time-nonlocality conditions  for a 
three level system as given in \cite{ali}: 
\begin{equation}
\label{prob3by2}
\left.
\begin{array}{lcl}
\rm prob(\hat{A}_1=+\frac{3}{2},\ \hat{B_1}=+\frac{3}{2})=0,\\
\rm prob(\hat{A}_1=+\frac{1}{2},\ \hat{B_2}=+\frac{3}{2})=0,\\
\rm prob(\hat{A}_1=-\frac{1}{2},\ \hat{B_2}=+\frac{3}{2})=0,\\
\rm prob(\hat{A}_1=-\frac{3}{2},\ \hat{B_2}=+\frac{3}{2})=0,\\
\rm prob(\hat{A}_2=+\frac{3}{2},\ \hat{B_1}=-\frac{3}{2})=0,\\
\rm prob(\hat{A}_2=+\frac{3}{2},\ \hat{B_1}=-\frac{1}{2})=0,\\
\rm prob(\hat{A}_2=+\frac{3}{2},\ \hat{B_1}=+\frac{1}{2})=0,\\
\rm prob(\hat{A}_2=+\frac{3}{2},\ \hat{B_2}=+\frac{3}{2})=q \ > 0.
\end{array}
\right \}
\end{equation}
The following setting achieves a maximal success probability
 1/4 of Hardy's argument:
\begin{eqnarray}\nonumber
 &|\psi\rangle=\sqrt{3}\tan\frac{\theta}{2}|\hat{S}_Z=+\frac{1}{2}\rangle+&\nonumber\\
&\sqrt{3}\tan^{2}\frac{\theta}{2}|\hat{S}_Z=-\frac{1}{2}\rangle+
\tan^{3}\frac{\theta}{2}|\hat{S}_Z=-\frac{3}{2}\rangle&\nonumber
\end{eqnarray}
\begin{equation}\nonumber
 \hat{A}_1=\hat{B}_2=\left( \begin{array}{cccc}
\frac{3}{2} &0 &0 &0\\
0 & \frac{1}{2} & 0 & 0\\
0 & 0 & -\frac{1}{2}& 0\\
 0 & 0 &0&-\frac{3}{2}\end{array} \right)=\hat{S}_Z
\end{equation}
\begin{equation}\nonumber
 \hat{A}_2=\hat{B}_1=\cos\theta\hat{S}_Z+\sin\theta\hat{S}_X
\end{equation}
 where
\begin{equation}\nonumber
 \hat{S}_X=\frac{1}{2}\left( \begin{array}{cccc}
0 &\sqrt{3} &0 &0\\
\sqrt{3} & 0 & 2 & 0\\
0 & 2 & 0& \sqrt{3}\\
 0 & 0 &\sqrt{3}&0\end{array} \right)
\end{equation}
and
\begin{equation}\label{cot}
 \cot^{6}\frac{\theta}{2}-3\cot^{4}\frac{\theta}{2}-3\cot^{2}\frac{\theta}{2}-1=0
\end{equation}
i.e, $\theta=2\cot^{-1}\frac{1}{\sqrt{1+2^{\frac{1}{3}}+2^{\frac{2}{3}}}}$.

The projectors $\Pi_{a_1,{A_{1}}},\Pi_{a_2,{A_{2}}}, \Pi_{b_1,{B_{1}}},\Pi_{b_2,{B_{2}}}$ 
in this case are given as 
\begin{eqnarray}\nonumber
 \Pi_{a_1,{A_{1}}}=|\hat{S}_Z=+\frac{3}{2}\rangle\langle\hat{S}_Z=+\frac{3}{2}|=\Pi_{b_2,{B_{2}}}\nonumber\\
\Pi_{a_2,{A_{2}}}=|\hat{A}_2=+\frac{3}{2}\rangle\langle\hat{A}_2=\frac{3}{2}|=\Pi_{b_1,{B_{1}}}\nonumber
\end{eqnarray}
where
\begin{eqnarray}\nonumber
 &|\hat{A}_2=+\frac{3}{2}\rangle=\frac{1}{\sqrt{2}}
|\hat{S}_Z=+\frac{3}{2}\rangle +\sqrt{\frac{3}{2}}\tan\frac{\theta}{2}|\hat{S}_Z=+
\frac{1}{2}\rangle +&\nonumber\\
&\sqrt{\frac{3}{2}}\tan^{2}\frac{\theta}{2}|\hat{S}_Z=-\frac{1}{2}\rangle+ 
\frac{1}{\sqrt{2}}\tan^{3}\frac{\theta}{2}|\hat{S}_Z=-\frac{3}{2}\rangle&\nonumber,
\end{eqnarray}
$\theta$ is given by equation (\ref{cot}).\\
 $\Pi_{+\frac{3}{2},{A_{2}}}|\psi\rangle$, in this case reads as
\begin{equation}\label{extra2}
 \Pi_{+\frac{3}{2},{A_{2}}}|\psi\rangle=\frac{1}{2}|\psi\rangle+\frac{1}{2}|\phi\rangle,
\end{equation}
where $|\phi\rangle=|\hat{S}_Z=+\frac{3}{2}\rangle$, which is orthogonal to $|\psi\rangle$.
 
\subsection*{General spin}
We conjecture that the maximum of the success probability appearing in the temporal version of Hardy's
argument is $25\%$ for any spin-$s$ system. We write below the state and observable setting which 
may achieve this maximum:
\begin{eqnarray}\nonumber
 \hat{A}_1=\hat{B}_2; \hat{A}_2=\hat{B}_1, |\psi\rangle\bot|\hat{A}_1=+s\rangle,\nonumber\\ 
\Pi_{+s,{A_{2}}}|\psi\rangle=\frac{1}{2}[|\psi\rangle+e^{i\eta}|\phi\rangle]\nonumber
\end{eqnarray}
where $|\phi\rangle\in\Pi_{+s,{B_{2}}}\mathcal{H}.$ 
\section*{Acknowledgement}
S.G. thankfully acknowledges the hospitality of the School of Chemistry and Physics, University
of KwaZulu-Natal during his visit to the school, during which a part of the work was done.


\begin{thebibliography}{9}
\bibitem{generalized} There is one more difference between \cite{legett} and \cite{brukner}: while in \cite{legett}, the
observer measures a single observable at different times under the supposition that its value is 
independent whether or not a measurement (of the same observable) was performed at some earlier or later time,
in \cite{brukner} at any given time the observer has a choice between two (or more) different measurement
settings. The value assigned to an observable $\hat{A}$ at time $t_1$ is independent of
whether or not a measurement was performed at some different time $t_0$ or
which observable at that time was measured.
\bibitem{tk} Ideal measurement here means a measurement with minimal state change, 
for example with respect to the fidelity as a measure of state change. Although not used in the 
assumptions of the original derivation of the temporal Bell inequality  \cite{brukner},
without the restriction to  ideal measurements, classical systems would violate the inequality too 
if invasive measurements are used. Hence the inequality could not be used to distinguish classical from quantum behavior.
\bibitem{bob} Alice and Bob may be one and the same observer, but the randomness in the choice of $\hat{A}_1$ 
or $\hat{A}_2$ and that of $\hat{B}_1$ or $\hat{B}_2$ must be independent.
\bibitem{sk} These conditions, in a different context,
were first introduced in \cite{kunkri} to show the spatial nonlocal 
feature of two spin-s systems.
\bibitem{foot} Without loss of generality, we assume that the state of the system
does not evolve with time between two successive measurements.
\end{thebibliography}
\end{document}